# The Tricritical Point of the *f*-electron Antiferromagnet USb$_2$ Driven by High Magnetic Fields

R. L. Stillwell,[1] I-L. Liu,[4,5,6] N. Harrison,[2,3] M. Jaime,[2,3] J. R. Jeffries,[1] and N. P. Butch[5,6]

[1]*Materials Science Division, Lawrence Livermore National Laboratory, Livermore, CA 94550*
[2]*High Magnetic Field Laboratory, Los Alamos National Laboratory, Los Alamos, NM 87545*
[3]*Materials Physics and Applications Division, Los Alamos National Laboratory, Los Alamos, NM 87545*
[4]*Chemical Physics Department, University of Maryland, College Park, MD 20742*
[5]*NIST Center for Neutron Research, National Institute of Science and Technology, Gaithersburg, MD 20899*
[6]*Center for Nanophysics and Advanced Materials, Department of Physics, University of Maryland, College Park, MD 20742*

ABSTRACT

In pulsed magnetic fields up to 65T and at temperatures below the Néel transition, our magnetization and magnetostriction measurements reveal a field-induced metamagnetic-like transition that is suggestive of an antiferromagnetic to polarized paramagnetic or ferrimagnetic ordering. Our data also suggests a change in the nature of this metamagnetic-like transition from second- to first-order-like near a tricritical point at $T_{tc}$ ~145K and $H_c$~52T. At high fields for H>$H_c$ we found a decreased magnetic moment roughly half of the moment reported in low field measurements. We propose that *f-p* hybridization effects and magnetoelastic interactions drive the decreased moment, lack of saturation at high fields, and the decreased phase boundary.

## I. Introduction

The rich physics of actinide and lanthanide materials provides a diverse palette to explore exotic phenomena driven by the itinerancy or localization of the *f*-electrons[1-3]. Uranium intermetallics in particular span the spectrum of materials properties from magnetism to superconductivity and hidden order to heavy fermions with strongly correlated 5*f*-electrons[3-6]. Since these materials are driven by such interactions they are particularly amenable to tuning via the application of pressure, doping, and magnetic fields [7-11]. Within uranium intermetallics, the uranium monopnictides and chalcogenides have garnered a great deal of attention because of their varied magnetic states supported by a simple crystal structure. These compounds variously order antiferromagnetically (UX, X=As, P, Sb), ferromagnetically (UY, Y=S, Se,Te), and ferrimagnetically (UAs, UP), sometimes within the same magnetic field or doping phase diagram. Within the UX and UY families the ordering temperatures can be tuned by varying the pnictogen or chalcogen anion. As a particular example, all of the uranium dipnictides ($UX_2$) transition from paramagnetic to antiferromagnetic (AFM) [12], yet while the Néel temperature, $T_N$, can be tuned by changing the pnictogen anion, neither $T_N$ nor the ordered magnetic moments change monotonically with increasing pnictogen radius[13]. Similar lattice tuning effects, in this case via high-pressure, revealed that $T_N$ in $USb_2$ is enhanced as a function of increasing pressure, yet at P= 9.8 GPa the AFM state unexpectedly disappears and is replaced by ferromagnetic ordering[9]. Given the success of tuning these compounds with doping and pressure, the question naturally arises as to whether they can also be tuned via the application of magnetic fields.

The use of magnetic fields has historically been limited to a select few of these compounds, since, commensurate with their higher ordering temperatures, the critical fields exceed those accessible by common, laboratory-based magnet systems. For those compounds that have been studied with magnetic fields, several of them have shown a field-induced metamagnetic-like transition, in which they are driven from one ordered state to another by the application of high magnetic fields[14-17]. Both $UBi_2$ and $USb_2$ have been shown to have some amount of hybridization between the *f*- and conduction electrons. This connection between the itinerant *f*-electrons and the local moments is seen by the Fermi surface reconstruction that occurs at the AFM transition. One of the other interesting differences within the $UX_2$ series is that $UBi_2$ is the only member that maintains the same Fermi surface structure when crossing from the paramagnetic (PM) phase into the AFM phase[7, 18, 19]. This is because of a lack of zonefolding due to the fact that the $UBi_2$ AFM state has a unit cell that is the same as the crystal structure. This difference in Fermi surfaces between UBi$_2$ and USb$_2$ is manifested most clearly in the anisotropy of the electrical transport along the c-axis (001), where there is a factor of three increase in the electrical resistivity upon entering the AFM state in USb$_2$, whereas there is a negligible increase for UBi$_2$[7, 20, 21]. This effect confirms that the Fermi surface reconstruction that occurs in USb$_2$ upon entering the AFM phase drives the anisotropic transport properties.

The Fermi surfaces of UBi$_2$ and USb$_2$ were measured by Aoki et *al.* deep in the ordered state of both materials (T~ 50mK). They determined that UBi$_2$ consists of one spherical Fermi surface at the center of the Brillouin zone and two cylindrical Fermi surfaces at the zone corners[7, 18]. In the paramagnetic state UBi$_2$ and USb$_2$ share this same Fermi surface structure but as USb$_2$ transitions into the AFM state the spherical surface at the zone center reconstructs into two cylindrical Fermi surfaces, and hence the high level of anisotropy along the c-axis. The effective masses for UBi$_2$ range from *m\**=4.4-9.2 *m$_0$* (where *m$_0$* is the bare electron mass) and the masses for USb$_2$ range from *m\**=2-6 *m$_0$*. These modestly high masses confirm that there is hybridization between the U 5*f* electrons and the conduction electrons. Taken together with electronic specific heat coefficients of ~20 mJ/mol-K$^2$ and the detection of a narrow dispersing band by angle-resolved photoemission spectroscopy, USb$_2$ is understood as a partially itinerant antiferromagnet with moderately strong electronic interactions[12, 22]. Given the hybridization in USb$_2$, the anisotropy of the magnetic ordering and high ordering temperature, it seemed reasonable to conclude that high-magnetic fields would be needed to perturb the ground state of this material and investigate the magnetic ordering.

High magnetic fields have proven to be a valuable tool to study many U based intermetallic and heavy fermion compounds because of their ability to reversibly tune and interrogate electronic properties and structure [4, 23]. Indeed, hybridization effects in many 5*f* materials, such as UNiAl and UCo$_2$Si$_2$, manifest as a field-induced metamagnetic-like transition at relatively high magnetic fields, followed by a slow saturation of the magnetic moment, an effect that occurs in fields in excess of 65 T in the case of UCo$_2$Si$_2$[24, 25]. With the development of facilities that can routinely reach 65 to 100 T non-destructively, we now have the ability to investigate many of these uranium compounds with ordering temperatures of $T_{N,C} \sim 100\text{-}200K$. To that end we have performed magnetostriction and magnetization measurements on USb$_2$ in high magnetic fields to address whether the AFM transition can be tuned with magnetic field and if there is a structural distortion concomitant with the magnetic transition.

## II. Experiment

Single crystals of USb$_2$ were grown via self flux with excess Sb using a U:Sb ratio of 1:6. Depleted U (3N7, New Brunswick Laboratories) and Sb (4N, ESPI Metals) were combined in an alumina crucible, which was sealed in a quartz tube under a partial pressure of UHP Ar. The materials were heated to 1100º C and held for 96 hours, then slow-cooled to 800º C over 100 hours, after which the excess flux was spun off in a centrifuge. The crystals formed as platelets up to about 5 mm on a side. Powder and Laue x-ray diffraction were used to confirm the crystal structure and single-crystal nature of the samples.

Magnetostriction and thermal expansion measurements were performed via an optical fiber Bragg grating (FBG) dilatometer[26, 27]. Oriented single-crystals of USb$_2$ were attached to a 125 μm diameter telecom-type optical fiber using cyanoacrylate adhesive. The samples were positioned along the length of the fiber where the index of refraction had been modulated so as to reflect a particular wavelength of light known as the Bragg wavelength ($\lambda_B$), which would shift due to mechanical compression or expansion of the fiber. In order to increase the signal from the magnetostriction along the c-axis, six single crystals (platelets ~0.2 mm thick) were stacked up and glued together to make a stack of ~1mm total length before attaching them to the fiber (Fig. 1, upper left inset). The reflected light was collected using a monochromator and an InGaAs line-array camera working at 47 kHz. The wavelength of the reflection peak as a function of time $\lambda_B(t)$ is used to compute the strain on the sample as a function of applied magnetic field, S(H), where S=ΔL/L. Magnetostriction measurements were made in pulsed magnetic fields up to 65 tesla at the National High Magnetic Field Laboratory (NHMFL), Los Alamos, with a pulse duration of 25 milliseconds. Sample temperature was controlled using a resistive heater on the probe and thermal connection to the sample in gas was made by attaching a 25μm gold wire between the sample and the Cernox thermometer using silver epoxy.

Magnetization measurements up to 65T were performed at the NHMFL using an inductively driven, compensated, extraction-coil magnetometer[28]. In addition to the compensated coils, electronic balancing was also used in order to compensate for any non-uniform thermal contraction of the coils during cooling. A 50x50x100μm rectangular sample was mounted in a PTFE capsule using Apiezon N grease with the [001] axis oriented along the applied magnetic field direction. The capsule was then mounted on a long, thin rod reaching from the sample capsule to the top of the probe that could extract the sample out of the coil in order to perform field sweeps of the empty coil for background subtraction. Additional magnetization measurements up to 15T were performed using the Vibrating Sample Magnetometer option of a Quantum Design Physical Properties Measurement System in order to calibrate the high field data collected at the NHMFL.

## III. Results and Discussion

### A. *Thermal Expansion and Magnetostriction*

Initial thermal expansion measurements were made in zero magnetic field where the AFM transition was visible as a second order transition in strain (S) versus temperature, and is clearly seen in the coefficient of the thermal expansion, $dS/dT$, or $\alpha(T)$, at $T_N$= 202.3K (bottom inset of Fig. 2), which matches with previously reported values[7, 13, 21, 29]. The USb$_2$ crystals form as platelets in the a-b plane, with the thinner layers forming along the c-axis. The thermal expansion data shown in Fig. 2 was acquired with the platelet oriented with the a-b plane along the fiber/magnetic field (Fig.1 upper inset). Magnetostriction experiments were performed on crystals oriented both with H along the a-b plane and along the c-axis.

The strain versus applied magnetic field with the field along the easy axis *c* shows a field-induced phase transition for T<$T_N$ (Fig. 3). The transition continues to move to higher applied fields as the temperature is lowered reaching $H_c$= 63T at T= 80K (Fig. 4). The magnetostriction for H along the a-b plane did not show the field-induced phase transition for any temperatures T<$T_N$ up to 65T, confirming that the magnetic hard axis is in the a-b plane and that there is a strong anisotropy between the c axis and the a-b plane in $USb_2$.

Interestingly, as the temperature is lowered the phase transition goes from second-order-like to first-order-like near T=150K. This can be seen clearly in figure 4, where the transition becomes nearly discontinuous for T≤ 122K. In addition to the crossover to first order, the amplitude of the transition also increases as the temperature is lowered (Figs. 4 and 5(a)). We determined the amplitude of the transition by fitting a straight line to the data for 10T before the transition and another straight line to the data for H>$H_c$, and then found the amplitude of the change between these two lines. This change in the lattice at the field-induced phase transition demonstrates that there is a strong magnetoelastic effect. The magnetoelastic coupling in $USb_2$ shown in our data is also seen in many other 5*f*-electron systems, including the uranium monochalcogenides, $UCu_{0.95}$Ge, and $UPt_2Si_2$ [8, 16, 30]. This magnetoelastic transition could indicate a structural phase transition accompanying the magnetic phase transition, similar to the cubic rock salt to rhombohedral distortion in uranium sulfide (US) as it enters the ferromagnetic state [31, 32]. Because neutron diffraction measurements at ambient pressure and zero magnetic field do not show such a structural transition in $USb_2$, this may represent a field-stabilized structural transition that is induced through field-enhanced magnetoelastic coupling.

### B. Magnetization

In order to further investigate the magnetoelastic transition we also performed magnetization measurements on a single crystal of $USb_2$ above and below the Neél transition. We clearly observed the field-induced metamagnetic-like transition for T<$T_N$ (Fig. 6), confirming our magnetostriction measurements. Surprisingly, the change of the magnetization at the transition is only ΔM~1$\mu_B$, nearly half of the expected moment of 1.88$\mu_B$/U found by inelastic neutron scattering and magnetization measurements [12, 29, 33]. We propose that the decreased moment is a result of *f-p* hybridization. The critical field for the transition increases with decreasing temperature, down to the lowest temperatures and highest fields studied in this experiment. The critical field for the transition is defined as the intersection point of two straight lines, one fit to the nearly vertical part of the data trace and the other fit to the lower field side of the trace leading up to the nearly vertical section (See appendix Fig. 1A). Error bars for the transition were calculated as the points at which the data trace deviated from the fit lines above and below the transition. There is very good agreement on $H_c$ from both the magnetization and the magnetostriction measurements.

Another consistent trend in the two measurements is that the transition grows increasingly sharp as the temperature is decreased. Indeed, the order of the transition seen in magnetization goes from second- to first-order-like near T= 150K, similar to our magnetostriction measurements. The M(H) curves for 155K<T<180K shown in Fig. 7 demonstrate that there is an initial upturn in the magnetization (H*) before the metamagnetic-like transition ($H_c$), whereas for T<155K there is no significant increase in magnetization preceding the transition.  To further investigate the connection between the metamagnetic-like transition and the tricritical point we plot the change in the transition widths of both H* and $H_c$ as a function of temperature (Fig. 8). There is a clear onset of the H* transition width for T>$T_{tc}$ that can be seen in both figures 7 and 8a. The transition width of the metamagnetic-like transition at $H_c$ also changes near the tricritical point but it is harder to define the endpoints since the transition is not as sharp at higher temperatures.  Taken together, the magnetostriction and magnetization as a function of magnetic field, and the transition widths as a function of temperature, give strong evidence for the presence of a tricritical point near T= 145K and H= 52T. This second-to-first-order transition is illustrated in the phase diagram in Fig. 9 by using a solid line along the phase boundary to represent the second-order like transitions and a dashed line along the phase boundary to represent the first-order-like transitions.

This second- to first-order-like tricritical point is similar to that seen in canonical metamagnets such as $FeCl_2$, $CoCl_2$, and $Ni(NO_3)_2 \cdot H_2O$ and more recently in $CeRh_2Si_2$ and other $(U,Ce)X_2Si_2$ systems [34-38]. Following the work of Bidaux et *al.* [39] we calculated the tricritical temperature using the mean-field formula for a two sublattice antiferromagnet with ferromagnetic intralayer coupling such that,

$$T_{tc} = \frac{2}{3} \frac{T_N(T_N + 2\theta)}{T_N + \theta}$$

where $T_{tc}$ is the tricritical point where the transition goes from second- to first-order like, $T_N$ is the Neél temperature, and $\theta$ is the Weiss constant, found from the slope of $1/\chi$ vs. T. Substituting $T_N$= 202K and $\theta$= +18K [29] into the equation above we calculate a tricritical temperature $T_{tc}$ = 145 K, corresponds to the temperature at which there is a change in transition amplitude for magnetostriction versus magnetic field (Fig. 5a).

To further characterize the high-field magnetic state for $H>H_c$, we also analyzed the slope of the strain versus magnetic field, shown in Fig. 5(b) and as an intensity plot in the high-field region of the phase diagram (Fig. 9). To determine the slope, we fit a straight line to all of the strain data for $H>H_c$ and then compared the slopes as a function of temperature, with the slope of the line represented by the color scale in Fig. 9 (see inset for scale values). The plot of the slope as a function of temperature in both Figs. 5(b) and 9, demonstrates that there is variation in slope above $T_{tc}$, but it then decreases and remains nearly zero for all $T<T_{tc}$. The saturation of the slope for $T<T_{tc}$ gives more evidence for a field-induced metamagnetic-like transition from the AFM state into a polarized paramagnetic or ferrimagnetic state. Although this AFM to ferrimagnetic transition is similar to other UX and UTX systems it is not possible to definitively claim what the order of the high field state is due to the small field range available above the transition as well as the noise in the high-field data [5, 14, 15, 40, 41].

# IV. Conclusions

Using high magnetic fields up to 65T we were able to see a field-induced metamagnetic-like transition in both magnetostriction and magnetization measurements and construct a *H-T* phase diagram for USb$_2$. We also observed that the order of the transition changes from second-order-like to first-order-like near a calculated tricritical point at T$_{tc}\sim$ 145K and H$_{tc}\sim$ 52T. By studying the slope of the magnetostriction above H$_c$ we were able to show evidence of a high field state suggestive of a polarized paramagnetic or ferrimagnetic order, though when considered with the jump in magnetization of $\sim$1$\mu_B$/U, it is more likely a ferrimagnetic state in which only half of the layers flipped. The combination of magnetic anisotropy and field-tuned magnetoelastic coupling suggest that future measurements using x-ray or neutron scattering at high field will yield important insight into details of the magnetic structure and *f*-electron hybridization.

This work was performed under LDRD (Tracking Code 14-ERD-041) and under the auspices of the US Department of Energy by Lawrence Livermore National Laboratory (LLNL) under Contract No. DE-AC52- 07NA27344.  A portion of this work was performed at the National High Magnetic Field Laboratory, which is supported by National Science Foundation Cooperative Agreement No. DMR-1157490, the State of Florida, and the U.S. Department of Energy.

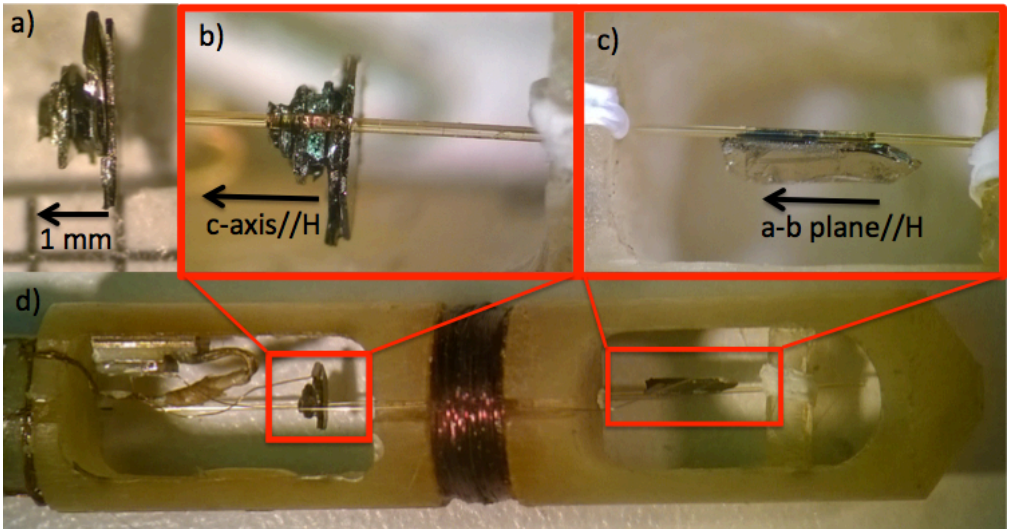

**FIG. 1.** (Color online) Experimental set-up for the fiber Bragg dilatometer showing the orientations of the USb$_2$ crystals. The samples shown in (a) were stacked along the c-axis to make a total height of 1 mm in order to increase the signal. (b) The stack was glued to the fiber optic cable so that the c-axis//H. (c) Crystal of USb$_2$ glued with the a-b plane of the crystal attached to the fiber. (d) Complete experimental setup showing the c-axis stack (lower left red box), the a-b plane attached along the crystal edge, the position of the thermometer (upper left corner), the magnetic pick-up coil used for magnetic field calibration wrapped around the holder (between the two samples), the gold wires attaching the samples to the thermometer in order to achieve thermal equilibrium.

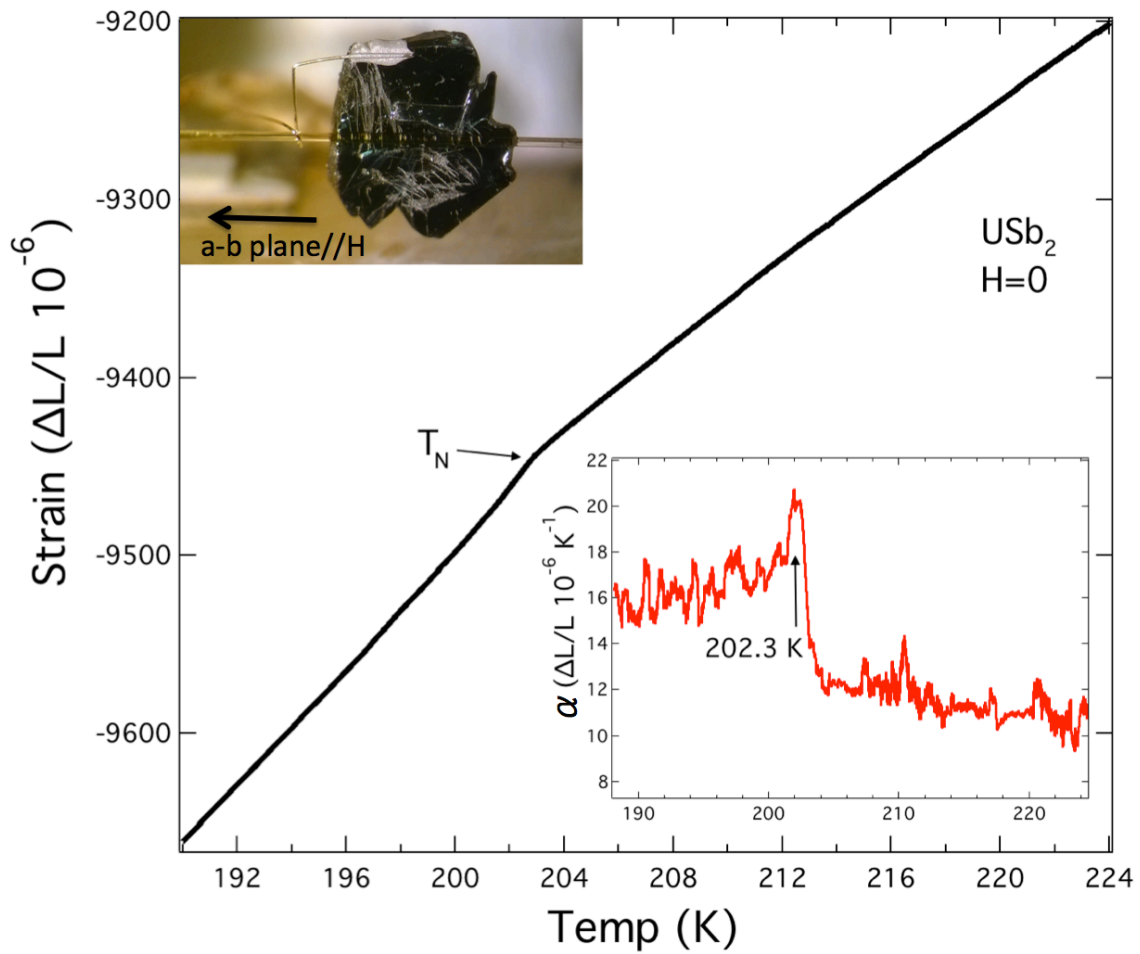

**FIG. 2.** (Color online) Thermal contraction as a function of temperature during cooling of USb$_2$ along the a-b plane. A picture of the samples attached to the fiber Bragg grating oriented with H along the a-b plane (shown in the top, right inset), and the stack of 6 crystals attached to the FBG with H along the c-axis (shown in the top, left inset). The coefficient of thermal contraction ($\alpha$) during cooldown (lower inset) is shown to emphasize the location of the AFM transition at T$_N$= 202.58 K, as previously reported in the literature.

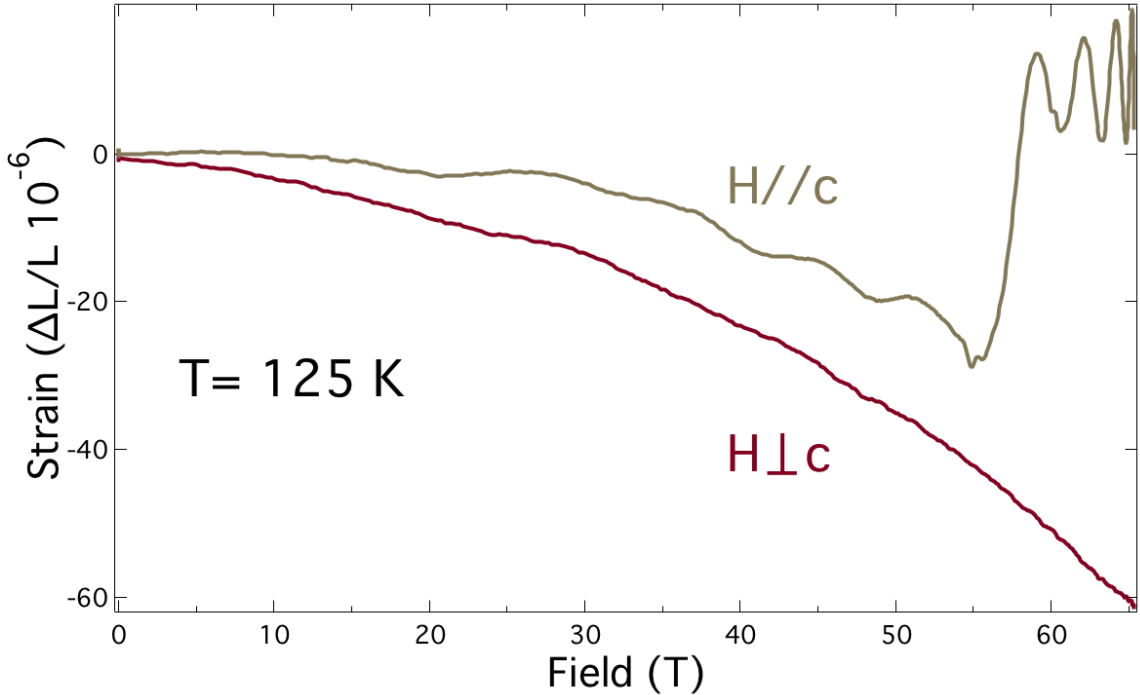

**FIG. 3.** (Color online) Comparison of the magnetostriction as a function of applied magnetic field, for H//c-axis and H perpendicular to the c-axis at T= 125 K. The anisotropy of the system is clearly demonstrated by the metamagnetic-like transition seen for H//c-axis, but not with H perpendicular to c-axis. All of the strain data shown in the following figures is for H//c-axis. Oscillations for H//c are likely due to vibrations resulting from the small misalignment of magnetic moment with the applied magnetic field.

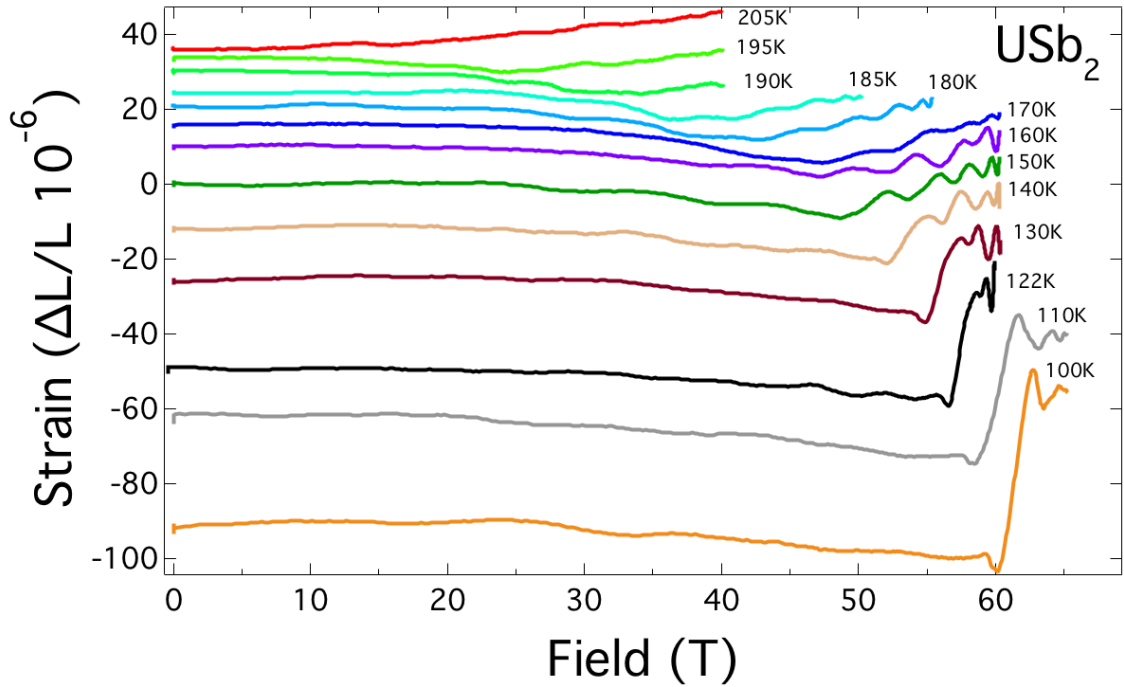

**FIG. 4.** (Color online) Strain as a function of applied magnetic field, measured with the **H**//c-axis. Measurements at temperatures above and below the AFM transition are shown. For T>$T_N$ there is no evidence of a transition,

but for T<$T_N$ there is a change in slope that grows increasingly sharp as the temperature is decreased, with a second-order to first-order like tricritical point at $T_{tc}$ ~ 145K. There is also a change in the slope of the strain for H>$H_c$ and T< 140K. All traces are offset vertically for clarity.

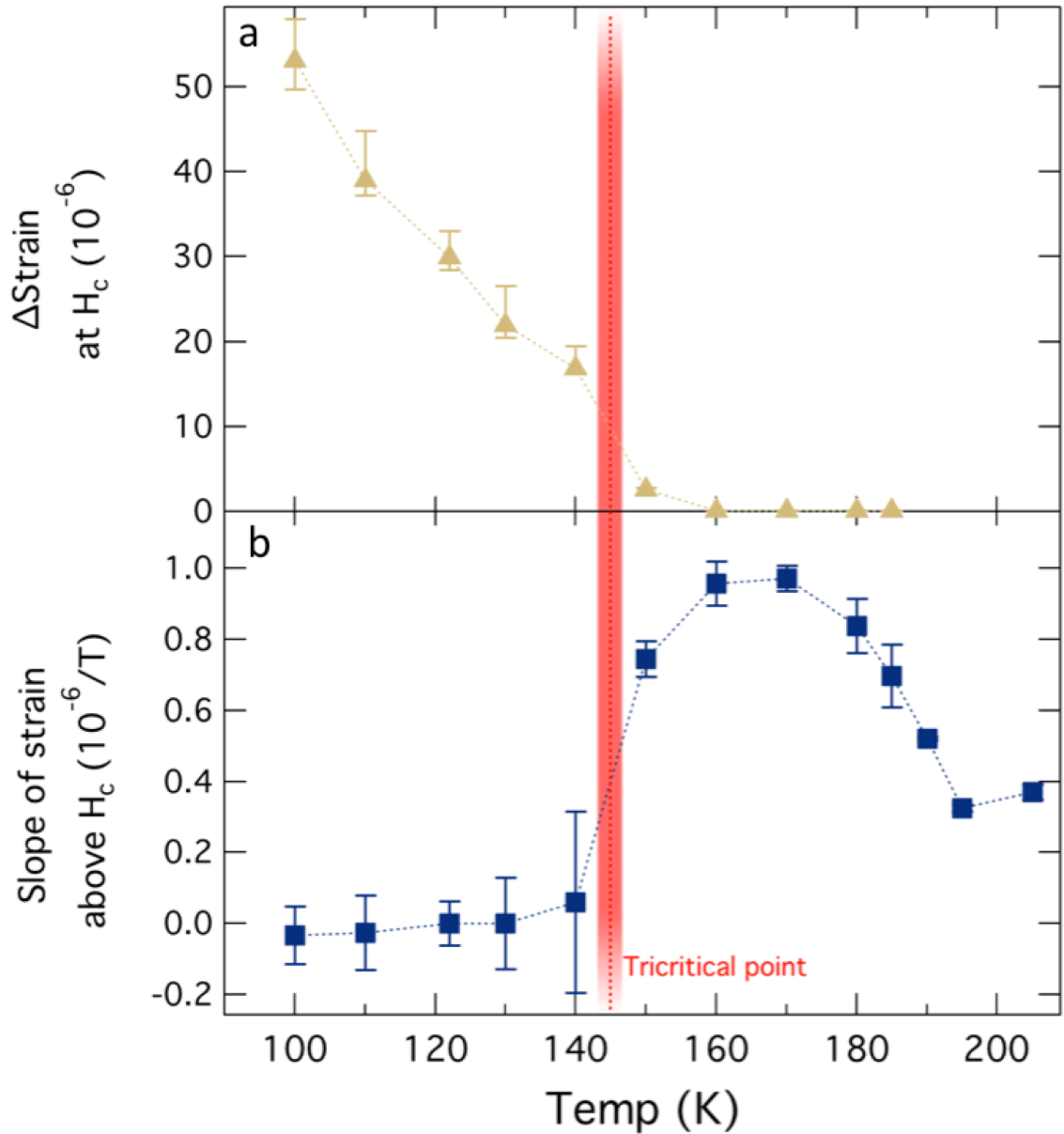

**FIG. 5.** (Color online) (a) Amplitude of the jump in strain at $H_c$ as a function of temperature. The red stripe at T= 145K is the tricritical temperature as predicted by theory. This data shows that there is a discontinuous jump in strain for T≤ 150K and that the amplitude of the jump at $H_c$ increases with decreasing temperature. (b) Plot of the slope of the strain above the transition (H>$H_c$) as a function of applied magnetic field. The slope of the strain at high temperatures increases initially with decreasing temperature and then decreases. For T≤ 140K the slope is nearly zero, within error, which we propose as evidence of a polarized paramagnetic or ferrimagnetic state..

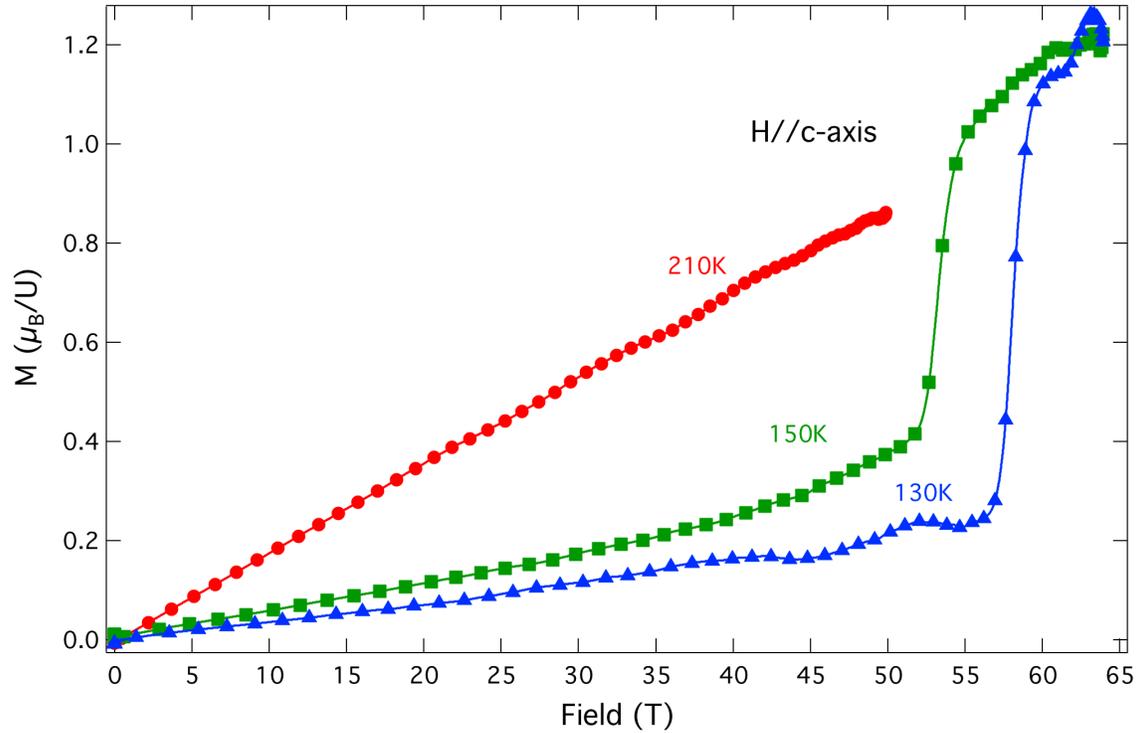

**FIG. 6.** (Color online) Magnetization as a function of applied magnetic field for H//c in units of $\mu_B$ per uranium atom for selected temperatures. For $T>T_N$ there is a linear trend of the magnetization but for $T<T_N$ there is a metamagnetic-like transition. The increase of the magnetization at $H_c$ is roughly $1\mu_B/U$, which is nearly a factor of two less than the previously measured zero-field moment of $1.88\mu_B/U$[29, 33]. High-field data was calibrated using a Quantum Design Vibrating Sample Magnetometer up to 15T.

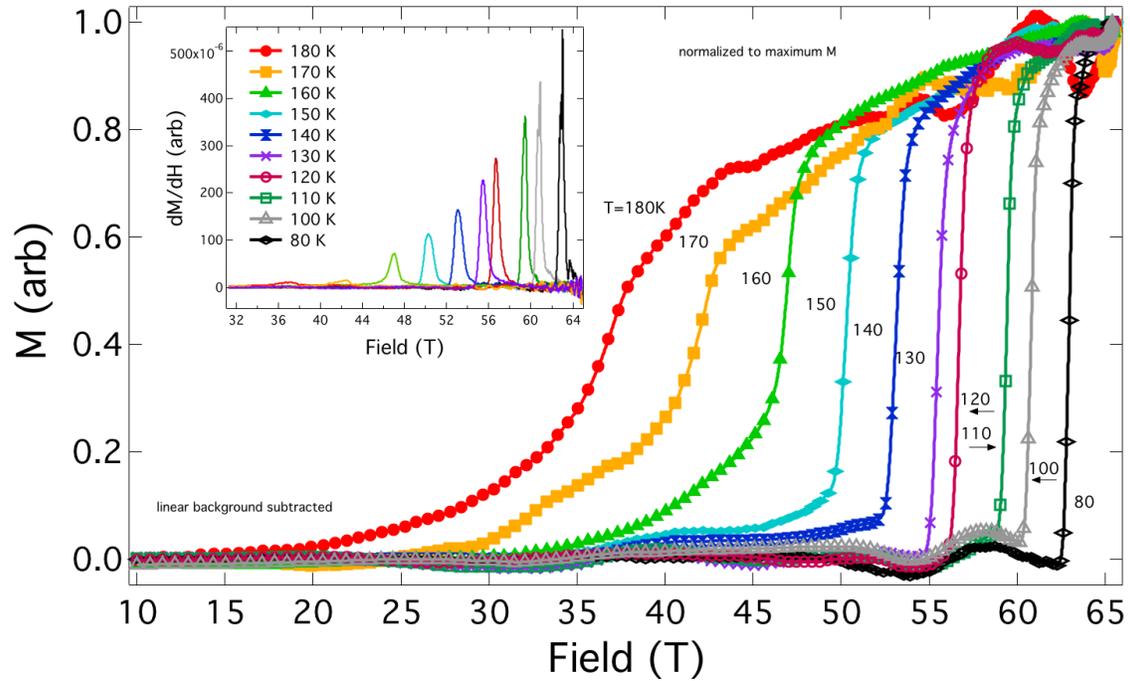

**FIG. 7.** (Color online) Plot of magnetization as a function of applied magnetic field for various temperatures. All of the traces were fit with a straight line from zero up to H*, where the slope of the trace deviates from linearity. All of the traces were then normalized so that the maximum value of the magnetization was equal to one. The upper inset of the figure shows the derivative of the magnetization as a function of magnetic field, dM/dH. The change from second to first order can be seen as the lower field side of the derivative becomes discontinuous for T<150K.

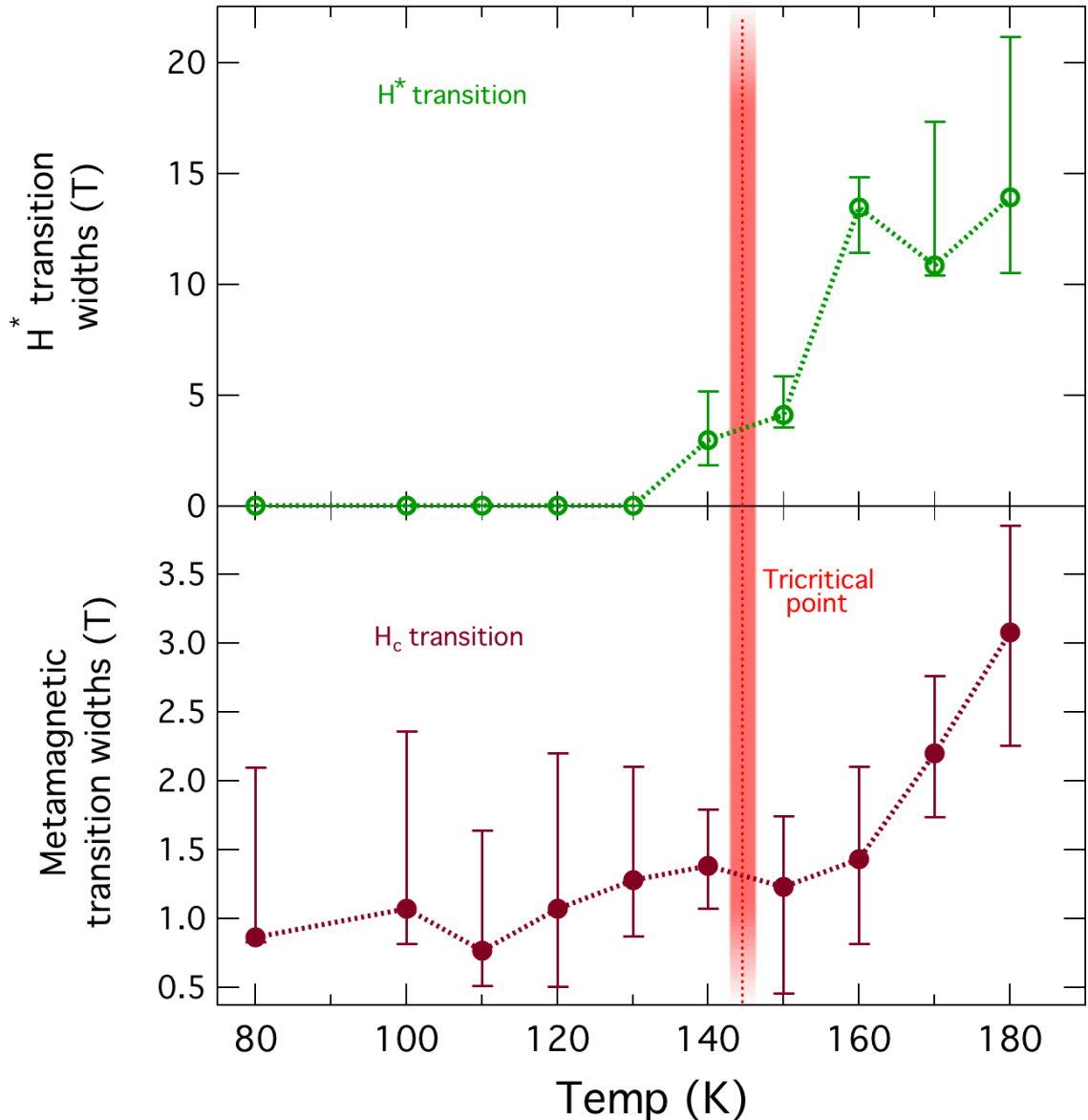

**FIG. 8.** (Color online) (a) Width of the second-order-like transition (H*) in the magnetization as a function of temperature. The width of the H* transition is clearly connected to the tricritical point at 145K, as the width of the transition goes to zero by 130K. (b) Plot of the width of the metamagnetic-liketransition, $H_c$, as a function of temperature. The $H_c$ transition width is not as strongly correlated as H* is to the order of the transition, nonetheless the trend of the data shows a decrease of the transition width as the temperature approaches $T \sim T_{tc}$. The error bars increase at low temperatures due to the narrow transition width and the noise above the transition.

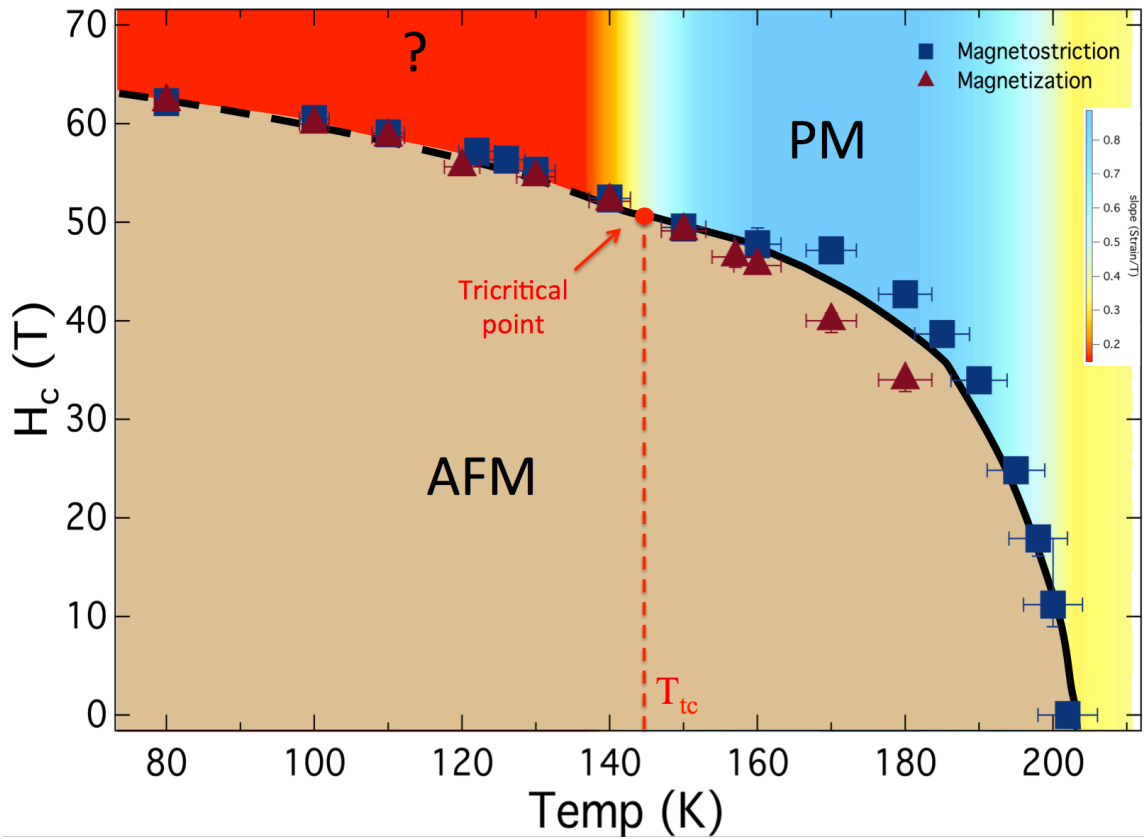

**FIG. 9.** (Color online) Magnetic field and temperature phase diagram of USb$_2$ (T$_N$=202.3K), compiled from both magnetostriction and magnetization data. The predicted tricritical point is shown at T$_{tc}$ =145K, with the red dashed line intersecting the phase boundary at H= 52T. The phase boundary is drawn with a solid black line for T> 145K to indicate that the metamagnetic-like transition is second-order like, whereas it is drawn as a dashed line for T< 145K where the metamagnetic-like transition is first-order like. The intensity plot above the phase boundary shows the slope of the strain versus magnetic field above the metamagnetic-like transition (H>H$_c$).

# Appendix

In order to better understand how the transitions in magnetization were chosen and how the errors were calculated, we have plotted two magnetization traces that are representative of a second order and a first order transition. Due to the second-order to first-order transition this was not a straightforward analysis, especially near the tricritical point. Figure 1A shows two magnetization traces, one at 160K where the transition is still second order, and another trace 80K where the transition is first order. The low field part of the curve was fit with a straight line, as was the curve above the transition. H* was chosen as the point where there was a clear deviation of the magnetization curve away from the straight line fit, indicated by the vertical dashed red line in the low-field side of the curve. The lower error in

H* was taken as the point where the straight line fit was no longer on top of the magnetization curve. The upper error in H* was taken as the point where the two straight-line fits intersect.

The transition for $H_c$ was taken as the onset of the jump in magnetization. The onset was found by doing a straight-line fit to the magnetization curve above and below the transition and then defining $H_c$ as the point where the two straight lines intersected. The upper and lower errors are taken as the points where the straight-line fits deviate from the magnetization curve, indicated in Fig. 1A with the dashed red circles. The transition width of H* is taken as the field range between the lower red, dashed straight line and $H_c$. The transition width of $H_c$ is taken as the field range between the red, dashed straight line at $H_c$ and the uppermost red, dashed straight line. The errors for the H* transition width is taken from the upper and lower errors for both transitions. The errors for the $H_c$ transition is taken from the transition error on the low-field side and the points where the magnetization curve deviates from the straight line fits on the high-field side of the curve.

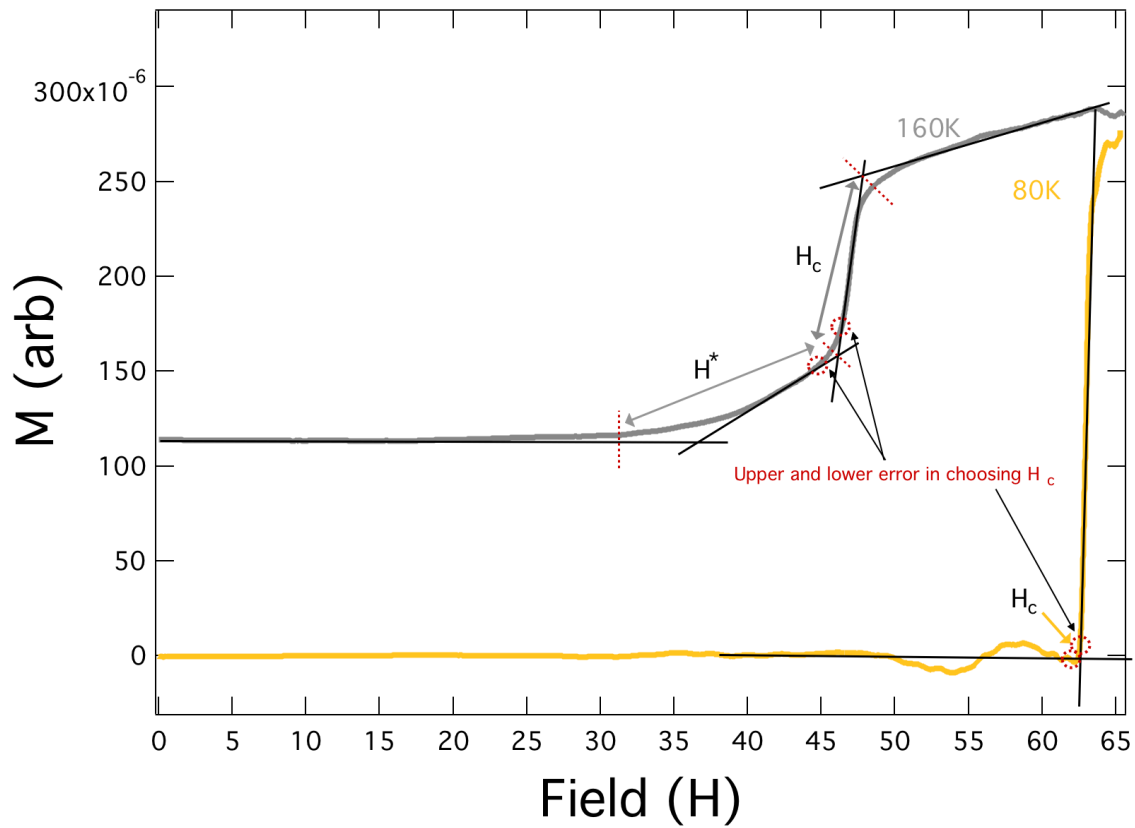

**FIG. 1A.** (Color online) Magnetization as a function of applied magnetic field for two temperatures showing how $H_c$, H* and the related error bars were chosen. $H_c$ is defined as the intersection point of two straight lines, one on the nearly vertical part of the trace and one on the part of the trace before going nearly vertical. The error is taken as the points where the trace deviates from the straight line. H* is taken as the point where there is a clear deviation of the magnetization curve from the straight-line fit.